\documentclass[3p,times,twocolumn]{elsarticle}

\usepackage{ecrcUpdate}

\volume{00}

\firstpage{1}

\journalname{\small Nuclear and Particle Physics Proceedings}

\runauth{David d'Enterria}

\jid{nppp}

\jnltitlelogo{\small Nucl. \& Part. Phys. Proceeds}

\usepackage{amssymb}

\usepackage{multirow}
\usepackage{xspace}

\newcommand{\sqrts}{\sqrt{s}}
\newcommand{\sqrtsnn}{\sqrt{s_{_{\mbox{\rm \tiny{NN}}}}}}

\newcommand{\pt}{p_{\rm T}}
\newcommand\cO{{\cal O}}

\newcommand{\LumiInt}{\mathcal{L}_{\mbox{\rm \tiny{int}}}}

\newcommand{\mcfm}{{\sc mcfm}}

\newcommand{\gaga}    {\gamma\,\gamma}
\newcommand{\qqbar}    {\rm q\,\bar{q}}
\newcommand{\bbbar}    {\rm b\,\bar{b}}
\newcommand{\ttbar}    {\rm t\,\bar{t}}

\newcommand{\MET}{\ensuremath{{E\!\!\!/}_{_{\rm T}}}}

\newcommand{\Lint}{\mathcal{L}_{\mbox{\rm \tiny{int}}}}

\newcommand*{\cm}{c.m.\@\xspace}

\def\cO#1{{{\cal{O}}}\left(#1\right)}
\def\ttt#1{\texttt{\small #1}}

\begin{document}

\begin{frontmatter}


\dochead{}

\title{Top-quark and Higgs boson perspectives at heavy-ion colliders}


\author{David d'Enterria}
\address{CERN, EP Department, 1211 Geneva, Switzerland}

\begin{abstract}
The perspectives for measuring the top quark and the Higgs boson in nuclear collisions at the LHC and 
Future Circular Collider (FCC) are summarized. Perturbative QCD calculations at (N)NLO accuracy, including 
nuclear parton distribution functions, are used to determine their cross sections and visible yields 
after standard analysis cuts in PbPb and pPb collisions at the LHC ($\sqrtsnn$~=~5.5,~8.8~TeV) and FCC 
($\sqrtsnn$~=~39,~63~TeV). In their ``cleanest'' decay channels, $\rm \ttbar\to \bbbar\,2\ell\,2\nu$ 
and $\rm H\to \gaga;\,4\ell$, about 10$^3$ ($10^5$) top-quark and 10 (10$^3$) Higgs-boson events
are expected at the LHC (FCC) for their total nominal integrated luminosities. Whereas the $\ttbar$ 
observation is clearcut at both colliders, evidence for Higgs production, perfectly possible at the 
FCC, requires integrating $\times$30 more luminosities at the LHC.
\end{abstract}


\end{frontmatter}


\section{Introduction}
\label{sec:intro}

The top quark and the Higgs boson (together with the $\tau$ lepton) are the only elementary Standard Model (SM)
particles that remain unobserved so far in nuclear collisions. Their production cross sections 
in hadronic collisions are dominated by gluon-gluon fusion processes ($\rm g\,g\to \ttbar+X;H+X$), 
computable today at the highest degree of theoretical accuracy: NNLO+NNLL for the top quark~\cite{Czakon:2013goa}, 
and N$^3$LO for the H boson~\cite{Anastasiou:2015ema}. The study of their yield modifications in heavy-ion compared to pp 
collisions would provide novel extremely well calibrated probes of the initial and final pA and AA states. 
Both elementary particles, the heaviest known, have very different decay channels and lifetimes, and can thereby be 
used to uniquely probe various aspects of strongly-interacting matter in nuclear collisions. On the one hand, the top-quark 
decays very rapidly before hadronizing ($\rm \tau_0 = \hbar/\Gamma_{t}\approx 0.1$~fm/c, much shorter than typical 
${\cal O}$(1~fm/c) QGP formation times) into $\rm t\to W\,b$ with $\sim$100\% branching ratio, with the W themselves 
decaying either leptonically ($\rm t\to W\,b \to \ell\,\nu,b$, 1/3 of the times) or hadronically ($\rm t\to W\,b \to \qqbar\,b$, 2/3 of the times). 
The kinematical distributions of the charged leptons ($\rm \ell = e,\mu$ unaffected by any final-state interactions) 
from $\ttbar\to \bbbar\,2\ell\,2\nu$ $\ttbar$ decays provide accurate information~\cite{d'Enterria:2015mgr} on the 
underlying nuclear gluon distribution function in the unexplored high-$x$ region where ``antishadowing'' and ``EMC'' 
effects are supposed to modify its shape compared to the free proton case~\cite{Eskola:2009uj}. On the other hand, 
the Higgs boson has a lifetime of $\rm \tau_0 \approx$~50~fm/c and, once produced, traverses the produced medium and 
scatters with the surrounding partons, resulting in a potential depletion of its yields compared to the pp case~\cite{DdECL}. 
The amount of Higgs boson suppression can be thereby used to accurately determine the final-state density of the produced QGP.
The perspectives of t-quark~\cite{d'Enterria:2015mgr} and Higgs boson~\cite{DdE} measurements in nuclear collisions
at current and future colliders are summarized here.


\begin{figure*}[hbpt!]
\centering
\includegraphics[width=0.4\textwidth,height=7.cm]{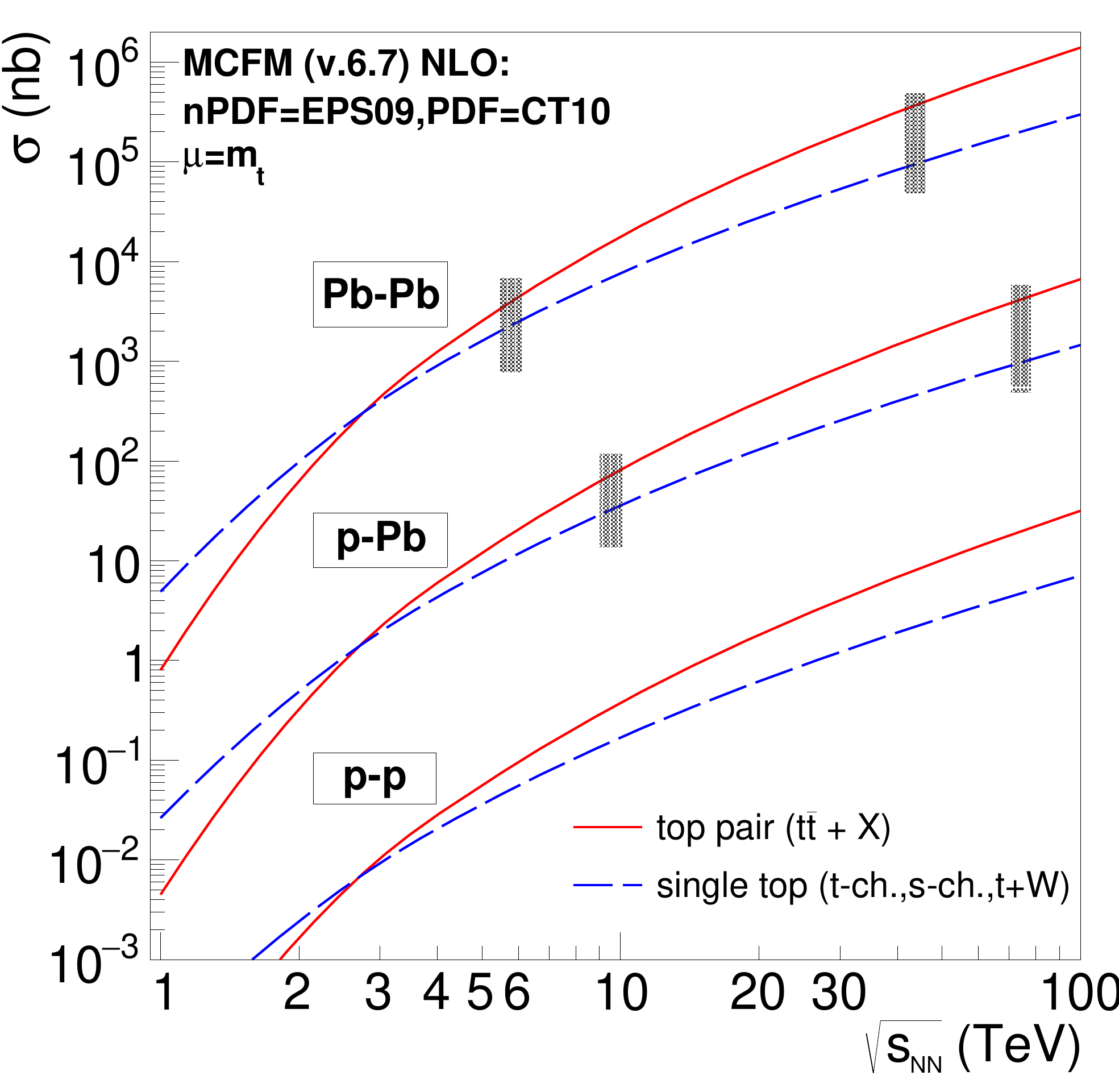}\hspace{0.5cm}
\includegraphics[width=0.4\textwidth,height=7.cm]{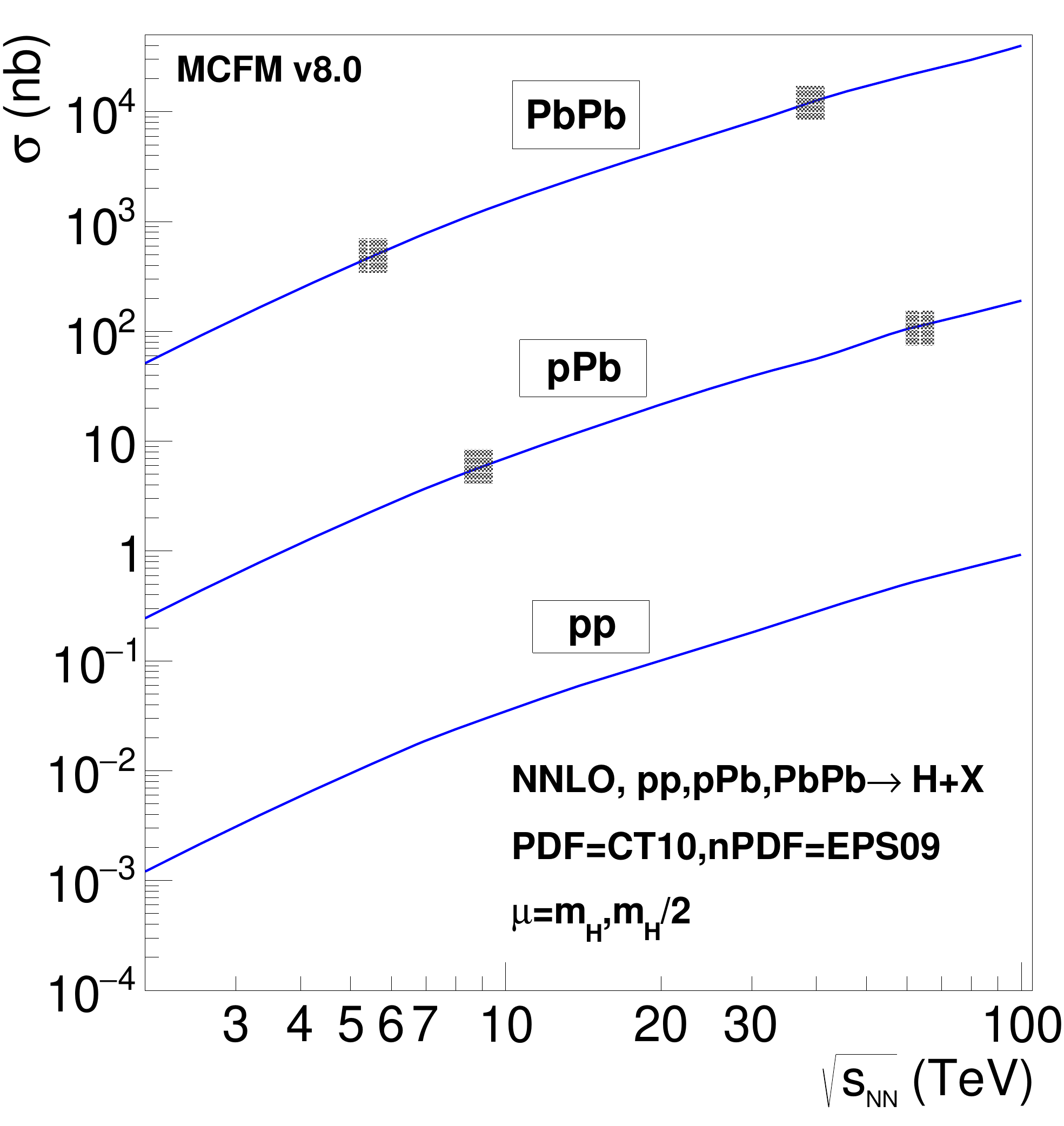}
\caption{Total production cross sections for top-pair and single-top at NLO (left)~\cite{d'Enterria:2015mgr} 
  and for the Higgs boson at NNLO (right)~\cite{DdE} in pp, pPb and PbPb collisions as a function of $\sqrtsnn$
  (the shaded boxes indicate the nominal LHC and FCC energies).}
\label{fig:xsec}
\end{figure*}

\begin{table*}[t]
\renewcommand\arraystretch{1.2}%
\caption{Cross sections, and expected number of counts after standard acceptance and efficiency cuts, 
  for $\ttbar$ and single-top (s-,t- and t\,W channels combined) at NLO (total, and in their respective leptonic decays)~\cite{d'Enterria:2015mgr} and Higgs boson 
  (total, and di-$\gamma$ and 4-lepton channels, at NNLO)~\cite{DdE} in pPb and PbPb collisions at LHC and FCC. 
  Maximum theoretical (scale and PDF) uncertainties (not quoted) are around 10\%. 
\label{tab:yields}}
\begin{center}
\begin{tabular}{lcc|cc|cc|ccc}\hline\hline
System & \hspace{-0.3cm} $\sqrtsnn$ \hspace{-0.2cm} & $\LumiInt$ & $\rm \ttbar$ & $\ttbar\to\bbbar\,\ell\ell\,\nu\nu$ & single-t & $\rm t\,W \to b\,\ell\ell\,\nu\nu$ & H & $\rm H\to \gaga$ & $\rm H\to Z\,Z^*(4\ell)$\\ 
      &  \hspace{-0.2cm}(TeV) \hspace{-0.2cm}  &            & $\sigma_{\rm tot}$ &  yields & $\sigma_{\rm tot}$ &  yields & $\sigma_{\rm tot}$ &  yields &  yields   \\ \hline
PbPb & \hspace{-0.2cm}5.5\hspace{-0.2cm}& 10 nb$^{-1}$ & 3.4 $\mu$b & 450  & 2.0 $\mu$b & 30 & 500 nb & 6 & 0.3 \\
pPb  & \hspace{-0.2cm}8.8\hspace{-0.2cm}& 1 pb$^{-1}$  & 59 nb      & 750 & 27 nb & 50 & 6.0 nb & 7 & 0.4 \\\hline
PbPb & \hspace{-0.2cm}39 \hspace{-0.2cm}& 33 nb$^{-1}$ & 300 $\mu$b & $1.5\times 10^5$ & 80 $\mu$b & 8000 & 11.5 $\mu$b & 450 & 25 \\
pPb  & \hspace{-0.2cm}63 \hspace{-0.2cm}& 8 pb$^{-1}$  & 3.2 $\mu$b & $4\times 10^5$& 775 nb & $2.1\times 10^4$ & 115 nb & 950 & 50 \\
\hline\hline
\end{tabular}
\end{center}
\end{table*}

The theoretical cross sections and yields in pPb and PbPb are obtained with \mcfm\ at NLO 
(v.6.7) for top-quarks~\cite{Campbell:2010ff}, and at NNLO (v.8.0) for the Higgs boson~\cite{Boughezal:2016wmq}, 
using CT10 proton PDFs~\cite{Lai:2010vv} and EPS09 nPDFs (including its 30 eigenvalues sets) for the Pb ion~\cite{Eskola:2009uj}. 
The following \mcfm\ processes are run: $\ttt{141}$ for $\ttbar$,  $\ttt{161,166,171,176,181,186}$ for single-top in the $t$-,$s$-channels
and associated with W~\cite{d'Enterria:2015mgr}; and $\ttt{119,116}$ for gluon-fusion $\rm H\to \gaga,\,4\ell$ (plus $\ttt{285,90}$ 
for the corresponding $\gaga,\,4\ell$ backgrounds), as well as $\ttt{91,101,215,540}$ corresponding to the rest of Higgs production 
channels (vector-boson-fusion, and associated with W, Z and top) to obtain the total H cross sections~\cite{DdE}.
All numerical results have been obtained with the latest SM parameters~\cite{PDG}, 
and fixing the default renormalization and factorization scales at 
$\rm \mu_F = \mu_R = m_{\rm t}$ for $\ttbar$ and single-top, 
$\rm \mu_F=\mu_R~= p_{_{\rm T,min;b-jet}}=50$~GeV for t\,W, and $\rm \mu_F=\mu_R=m_{\rm H}/2$ for Higgs.
These calculations reproduce very well the top and Higgs cross sections measured 
in pp collisions at $\sqrts$~=~7,~8,~13~TeV at the LHC. 
The collision energy dependence of the total top-quark and Higgs boson cross sections are shown in Fig.~\ref{fig:xsec}.
The cross sections increase by a factor of $\times$55--90 for $\ttbar$ and $\times$20 for the Higgs boson
between LHC and FCC energies.
The cross sections at the nominal LHC and FCC energies are listed in Table~\ref{tab:yields}.
Compared to the corresponding pp results at each \cm\ energy, antishadowing nPDF modifications increase 
the total top-quark yields by 2--8\%, whereas the Higgs cross sections are just slightly enhanced (depleted)
by $\sim$3\% at the LHC (FCC).
The PDF uncertainties, obtained adding in quadrature the CT10 and EPS09 uncertainties, are around (below) 10\%
for the top (Higgs) case. The $\cO{5-10\%}$ theoretical $\mu_F, \mu_R$ scale uncertainties, not quoted either, 
cancel out in the 
ratios of (pPb,PbPb)/(pp) cross sections at the same \cm\ energy.

\vspace{-0.2cm}

\section{Top quark measurement}
\label{sec:top}

\begin{figure*}[t]
\centering
\includegraphics[width=0.47\textwidth,height=5.cm]{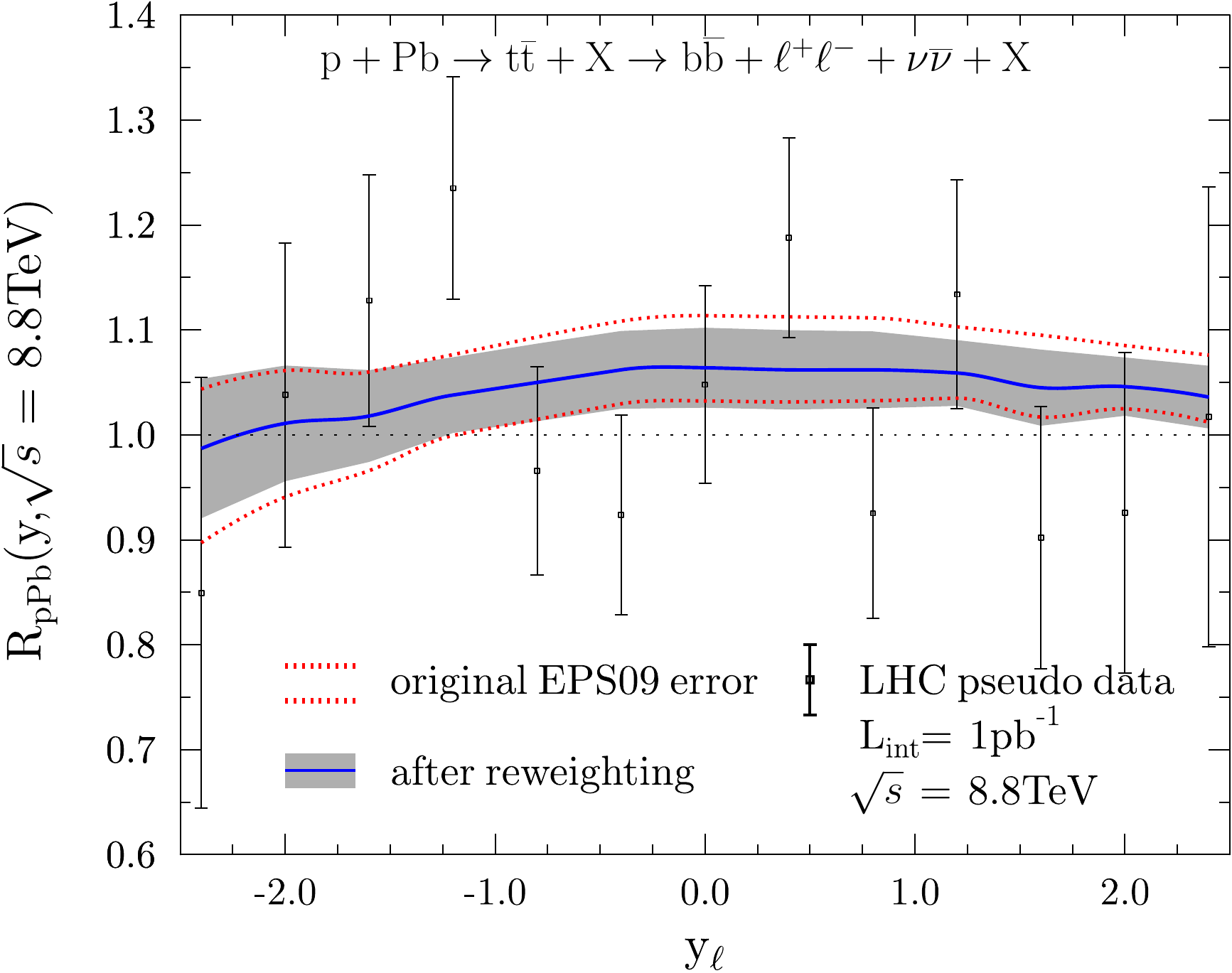}
\includegraphics[width=0.47\textwidth,height=5.cm]{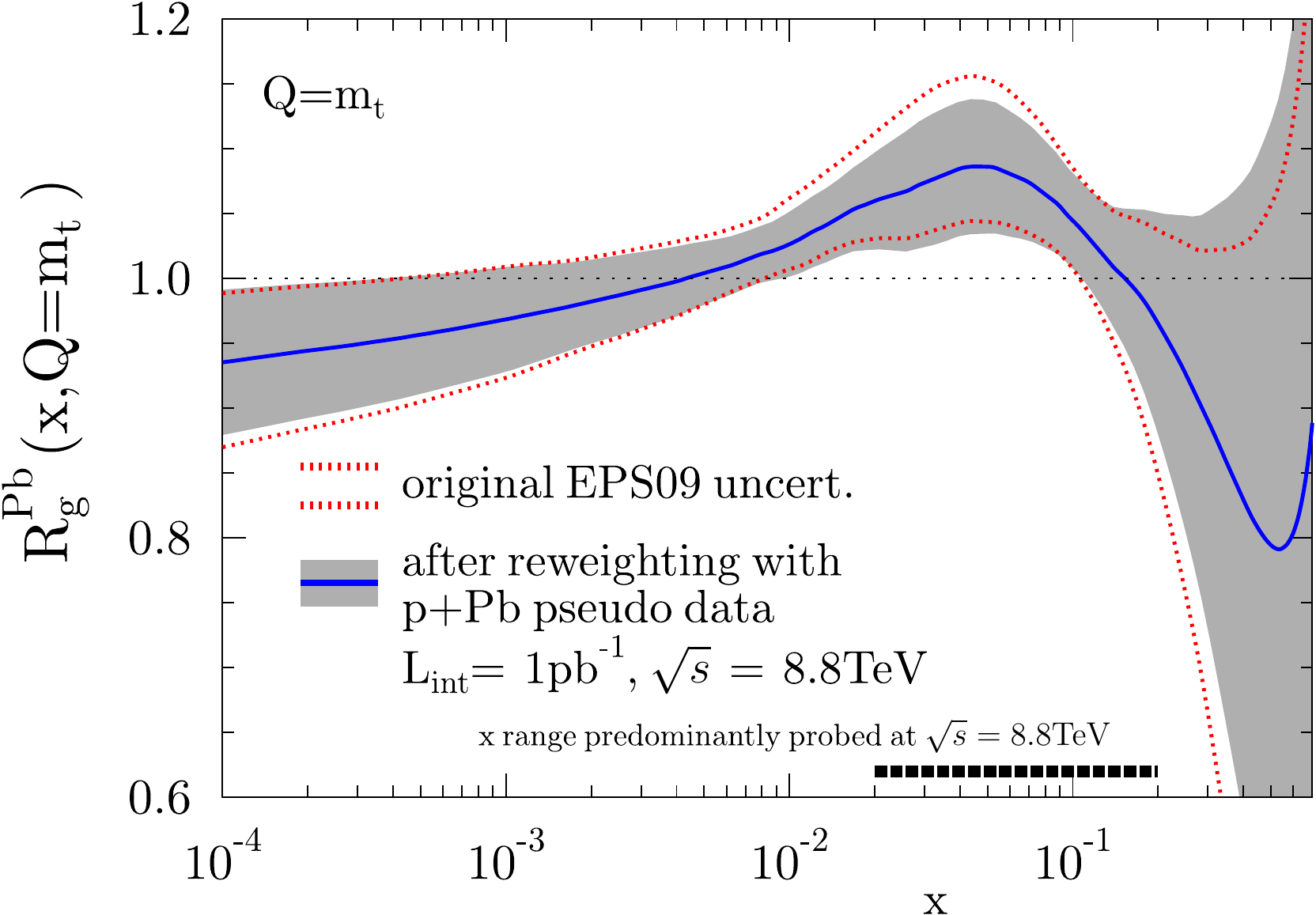}
\includegraphics[width=0.47\textwidth,height=5.cm]{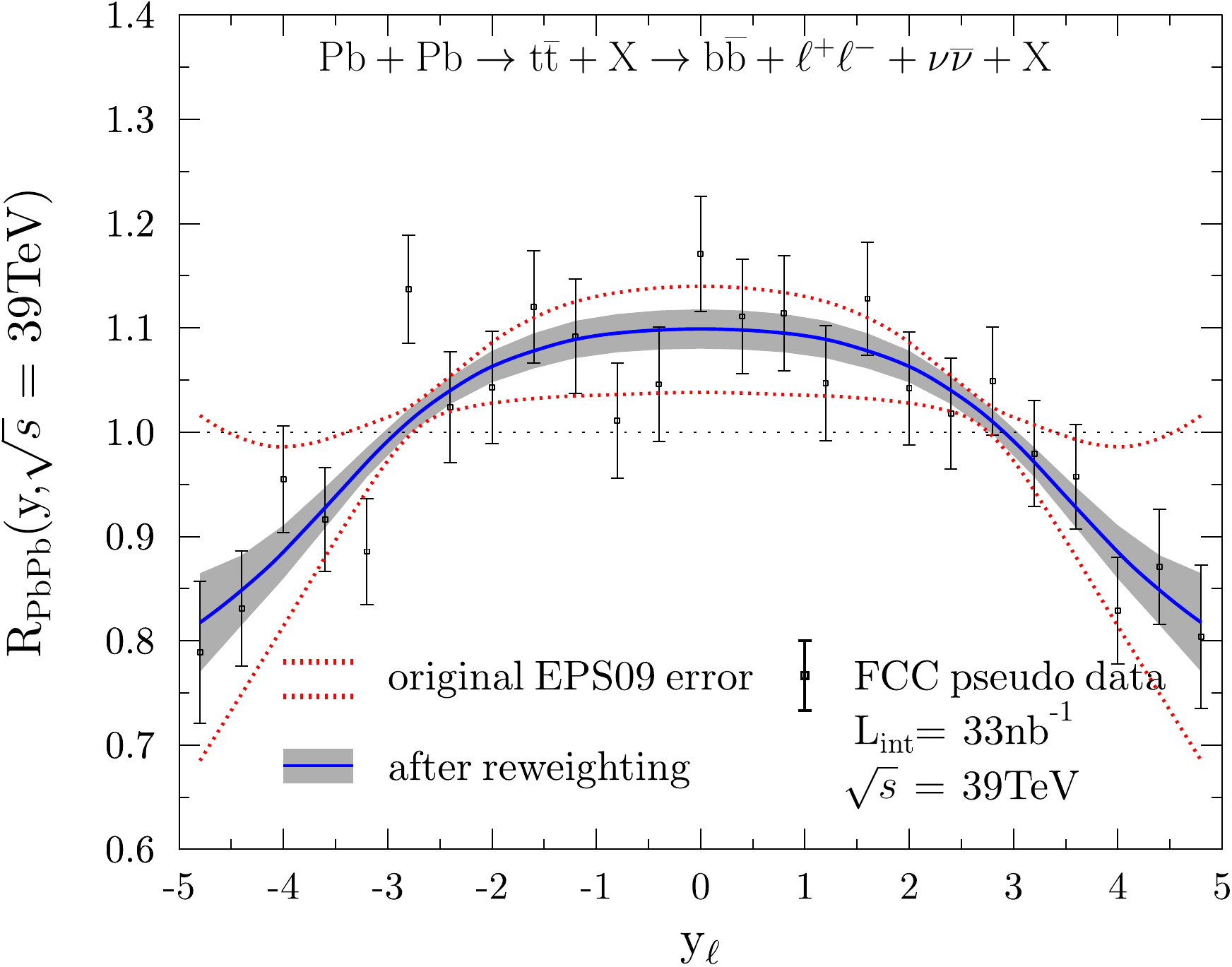}
\includegraphics[width=0.47\textwidth,height=5.cm]{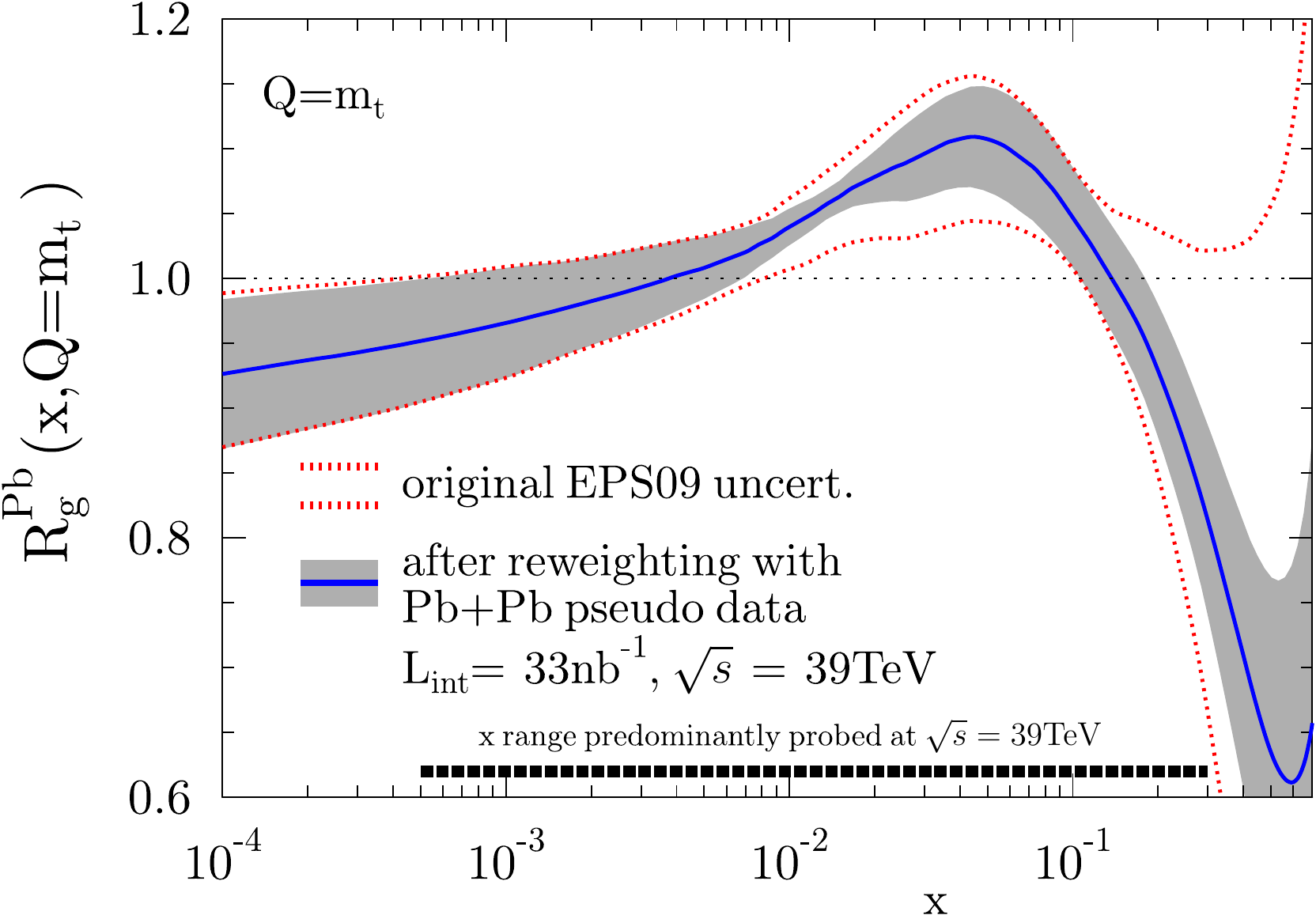}
\caption{Left: Pseudodata for nuclear modification factors as a function of rapidity of charged leptons 
produced in $\ttbar$ decays at the LHC (pPb, top)~\cite{d'Enterria:2015mgr} and FCC (PbPb, bottom)~\cite{Dainese:2016gch}.
Right: Original EPS09 gluon nuclear modification at $Q=m_{\rm top}$ (red dashed curves) and estimated improvements 
obtained by Hessian reweighting (grey area) using the LHC (pPb, top)~\cite{d'Enterria:2015mgr} and FCC (PbPb, bottom)~\cite{Dainese:2016gch} 
pseudodata.} \label{fig:ttbar} 
\end{figure*}

Top quarks decay almost exclusively to a b quark and a W boson and, in a heavy-ion environment, it is the
W leptonic decays that can be best resolved from the backgrounds. The estimated measurable yields using
realistic luminosities and analysis cuts (b-jets: anti-$k_{\rm T}$ algorithm with $R=0.5$, 
$\pt > 30$~GeV/c, $|\eta| < 3,\,5$\,(LHC,\,FCC), 50\% b-tagging efficiency; charged leptons: $R_{\rm isol} = 0.3$, 
$\pt > 20$~GeV/c, $|\eta| < 3,\,5$\,(LHC,\,FCC), $\rm m_{\ell\ell}>20$~GeV, $\rm |m_{\ell\ell}-m_Z|>15$~GeV; 
neutrinos: missing transverse energy $\MET > 40$~GeV) are listed in Table~\ref{tab:yields}. 
After branching ratios (BR~$\approx$~5\% for $\ttbar$, 22\% for single-t) and
acceptance and efficiency losses (40--50\% for $\ttbar$, 20--30\% for single-t), we expect
about 500 and 800 leptonic top-quark events in PbPb and pPb collisions at the LHC (with 
controllable backgrounds in the $\ttbar$ case). At FCC, the higher \cm\ energy and luminosities
will yield very large data samples with 150- and 400-thousand pairs in PbPb and pPb.
The resulting nuclear modification ratios of $\ttbar$ decay leptons yields as a function of 
rapidity, $R_{\rm pPb}(y_\ell) = d\sigma_{\rm pPb}(y_\ell)/(A\,d\sigma_{\rm pp}(y_\ell))$ and 
$R_{\rm PbPb}(y_\ell) = d\sigma_{\rm PbPb}(y_\ell)/(A^2\,d\sigma_{\rm pp}(y_\ell))$,
are shown in Fig.~\ref{fig:ttbar} (left) for pPb at the LHC 
(top) and PbPb at the FCC (bottom). The effect of antishadowing (shadowing) in the nPDF results 
in small enhancements (deficits) at central (forward) rapidities $y_\ell\approx$~0 ($|y_\ell|\gtrsim$~3).
The impact that these pseudodata would have in an EPS09 global fit of the nuclear PDFs is 
quantified via the Hessian reweighting technique~\cite{Paukkunen:2014zia}. 
The expected improvements of the gluon PDF nuclear modification factor
$R_{\rm g}^{\rm Pb}(x,Q^2) = g^{\rm Pb}(x,Q^2)/g^{\rm p}(x,Q^2)$ at $Q^2=m^2_{\rm top}$ 
are shown in Fig.~\ref{fig:ttbar} (right). The uncertainties on the nuclear 
gluon PDF are reduced by more than 10\% (50\%) at the LHC (FCC) 
mostly in the antishadowing and EMC $x$ regions~\cite{d'Enterria:2015mgr,Dainese:2016gch}.

\vspace{-0.2cm}

\section{Higgs boson measurement}
\label{sec:higgs}

\begin{figure*}[t]
\centering
\includegraphics[width=0.45\textwidth,height=6.1cm]{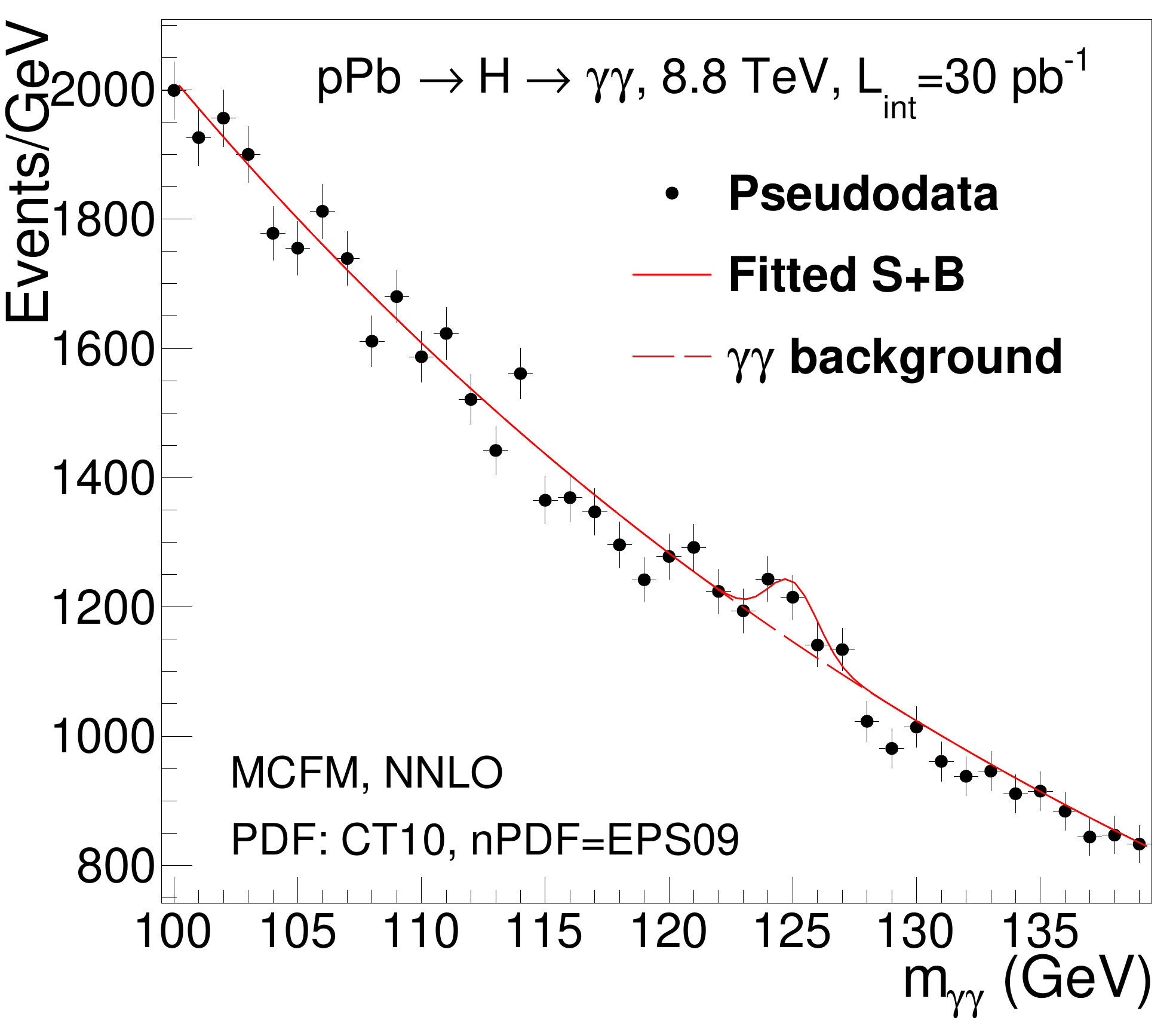}\hspace{0.5cm}
\includegraphics[width=0.45\textwidth,height=6.1cm]{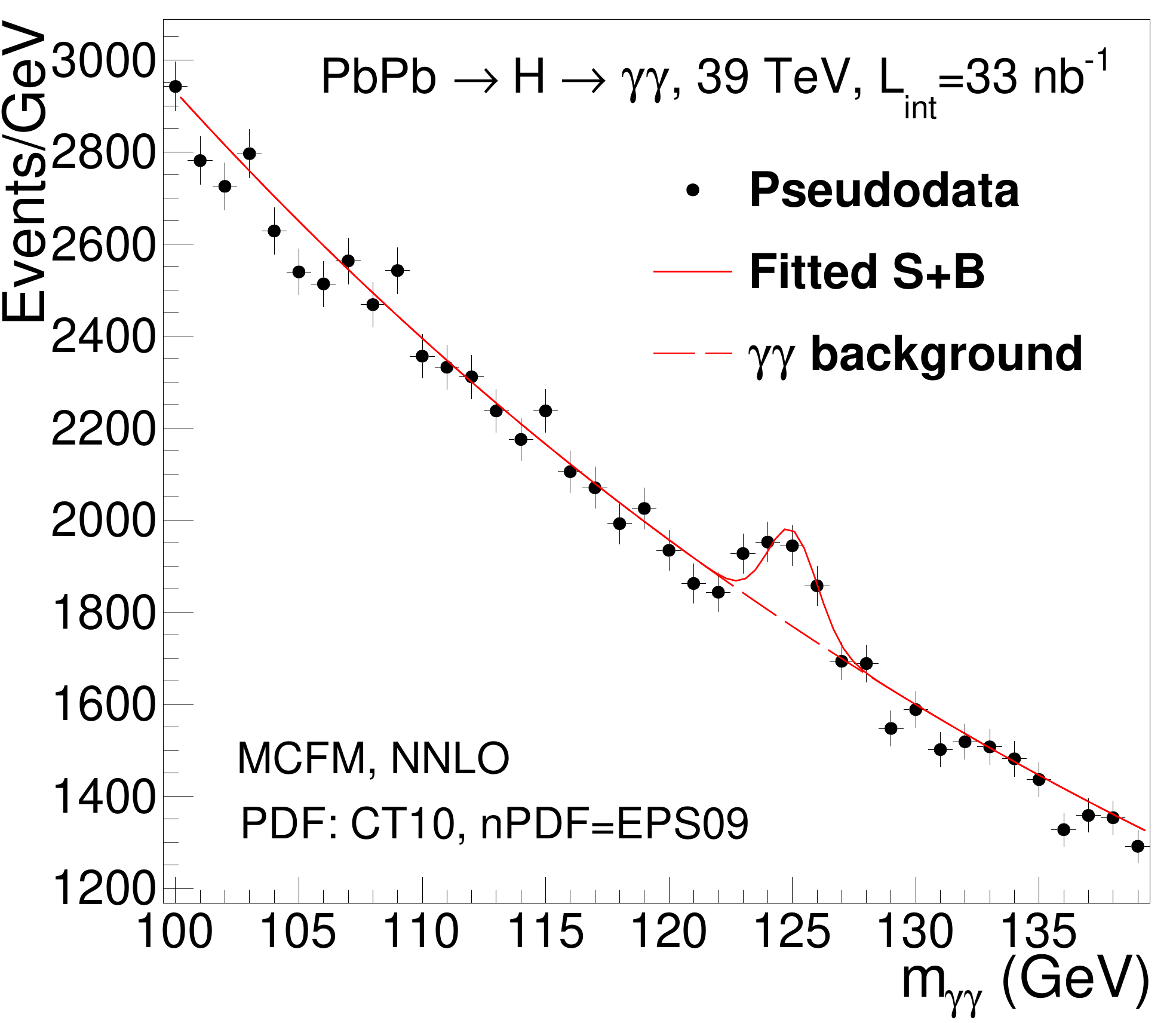}
\caption{Expected invariant mass diphoton distributions in pPb at $\sqrtsnn$~=~8.8~TeV (with enhanced LHC 
integrated luminosities, $\Lint$~=~30~pb$^{-1}$, left) and in PbPb at $\sqrtsnn$~=~39~TeV (nominal 
FCC luminosities, right) for a Higgs boson signal injected on top of the $\gaga$ backgrounds~\cite{DdE}.} 
\label{fig:Higgs}
\end{figure*}

The discovery of the Higgs boson in pp collisions at the LHC was carried out in the
clean diphoton and four-lepton channels with very low branching ratios but small and/or controllable
backgrounds~\cite{Aad:2012,Chatrchyan:2012xdj}. A first measurement in heavy-ion collisions
will certainly exploit the same final states. Table~\ref{tab:yields} lists the estimated measurable 
yields for nominal luminosities after accounting for typical ATLAS/CMS analysis cuts (photons: 
$\pt > 30,40$~GeV/c, $|\eta| < 2.5,\,5$\,(LHC,\,FCC), $R_{\rm isol} = 0.3$; charged leptons: 
$\pt > 20,15,10,10$~GeV/c, $|\eta| < 2.5,\,5$\,(LHC,\,FCC), $R_{\rm isol} = 0.3$). 
After branching ratios (BR~=~0.23\% for $\gaga$, 0.012\% for $4\,\ell$) and signal losses from 
acceptance and efficiency (45--60\% for diphotons, 50\% for 4-leptons), we expect about 10 
(1\,000) Higgs bosons in PbPb and pPb collisions at the LHC (FCC), on top of the corresponding 
$\gaga$ and $4\ell$ non-resonant backgrounds~\cite{DdE}. 
In the $\gaga$ case, the backgrounds include the irreducible QCD diphoton continuum plus
30\% of events coming from misidentified $\gamma$-jet and jet-jet processes.
For the nominal LHC luminosities, the significance of the diphoton signal (S) over the 
background (B), computed via S/$\sqrt{\rm B}$ at the Higgs peak,
is 0.5,0.6$\sigma$ in PbPb and pPb collisions, and thus one needs a factor of 
$\times$30--40 larger integrated luminosities to reach $3\sigma$ evidence (which would 
be further enhanced by including the $\rm H \to 4\ell$ results). Such a large $\Lint$ increase
would require, first, running one full-year (instead of the nominal heavy-ion month)
plus increasing the instantaneous luminosity by a factor of 3--4. Whereas the feasibility 
of such a setup seems difficult in the PbPb case, it's not unrealistic in the pPb mode
during the high-luminosity LHC phase (HL-LHC).
Figure~\ref{fig:Higgs} shows the expected diphoton invariant mass distributions for
pPb at the LHC (with $\Lint = 30$~pb$^{-1}$, left) and in PbPb collisions 
at the FCC. The significance of the H$(\gaga)$ peaks are 3$\sigma$ and 5.5$\sigma$
respectively.

\vspace{0.2cm}
\noindent {\bf Acknowledgments--} Discussions and/or common work with A.~David, C.~Loizides, 
and H.~Paukkunen are gratefully acknowledged.









\end{document}